\documentstyle[preprint,aps,floats,psfig]{revtex}
\begin{document}

\title  {Tunneling spectra for ($d_{x^2-y^2}+is$)-wave superconductors
versus tunneling spectra for ($d_{x^2-y^2}+id_{xy}$)-wave superconductors}

\author{ N. Stefanakis}
\address{ Department of Physics, University of Crete,
	P.O. Box 2208, GR-71003, Heraklion, Crete, Greece}

\date{\today}
 
\maketitle

\begin{abstract}
The tunneling conductance spectra of a normal metal / insulator / singlet superconductor is calculated from the reflection amplitudes using the 
Blonder-Tinkham-Klapwijk (BTK) formulation. The pairing symmetry of the 
superconductor is assumed to be $d_{x^2-y^2}+is$, or $d_{x^2-y^2}+id_{xy}$. 
It is found that in the ($d_{x^2-y^2}+is$)-wave 
case there is a well defined conductance peak in the 
conductance spectra, in the amplitude of the secondary $s$-wave component.
In the ($d_{x^2-y^2}+id_{xy}$)-wave case the tunneling 
conductance has residual values within the gap, due to the formation 
of bound states. The bound state energies depend on the angle of the 
incident quasiparticles, and also on the boundary orientation. 
On the basis of this observation an electron focusing experiment is proposed 
to probe the ($d_{x^2-y^2}+id_{xy}$)-wave state.
\end{abstract}


\section{Introduction}
Two decades ago, Blonder, et. al. \cite{blonder} used the Bogoliubov-de
Gennes (BdG) equations to calculate the 
tunneling conductance of normal metal /
s-wave superconductor contacts, with a barrier of arbitrary strength 
between them, in terms of the 
probability amplitudes of Andreev \cite{andreev} and normal 
reflection.
In the Andreev reflection process an electron incident, 
in the barrier can be reflected as an electron (normal reflection), 
reflected as a hole without changing its momentum (Andreev reflection),
it can also be transmitted into the superconductor as an electron-like, 
hole-like quasiparticle. 

Recently the BTK theory was extended by several groups to consider the 
anisotropy of the pair potential. 
In $d$-wave superconductors the pair potential changes sign under a
$90^o$-rotation. So under appropriate orientation of the $a$-axis
of $d$-wave superconductor the transmitted quasiparticle feel 
different sign of the pair potential. This results in the formation 
of bound states within the energy gap, which are detected as peaks in the 
conductance spectra. In $d$-wave superconductor the peak exists at 
$E=0$ for a great range of angles of incidence of the incoming 
electron. This range depends on the 
surface orientation. \cite{tanaka1}
In particular for (110) surfaces the peak exists at $E=0$ for all
angles of incidence, and disappears for the (010) or (100) surface.

In the presence of another barrier inside the normal metal 
additional subgap bound states exist due to multiple
Andreev reflections \cite{tanaka3,ting1}. 
The same phenomenon occurs in 
$d$-wave superconductor /
insulator / $d$-wave superconductor. In these systems the 
quasiparticle current has been examined by several 
groups \cite{averin,hurd,yoshida}, using BTK formalism with 
recursive relations for the determination of the probability 
amplitudes.

There is a competition between different pairing symmetries in the bulk. 
The coexistence of a subdominant order parameter in bulk depends 
on the strength of the secondary order parameter attractive interaction 
relative to the attractive interaction in the dominant pairing channel.
When the secondary order parameter is strong enough a second 
phase transition occurs at a temperature $T_{c1} < T_c$ which 
depends on the strength of the secondary order parameter.
Numerical results show that when such coexistence is realized the 
relative phase of the order parameters is $\pi /2$ leading to 
$d_{x^2-y^2}+is$ or $d_{x^2-y^2}+id_{xy}$ pairing state in the bulk. 
The temperature dependence of the various thermodynamic quantities and
transport properties change from power lows to exponential below $T_{c1}$.
\cite{ghosh1,ghosh2}
When the secondary order parameter is not strong enough, only the 
$d_{x^2-y^2}$-wave order parameter appears in the bulk. 
For $(110)$ surfaces the $d_{x^2-y^2}$-wave order parameter changes 
sign under reflection at the surface and vanishes at the surface. 
On the other hand the $s$ or $d_{xy}$ does not change sign and are
not effected by the presence of the surface, so there is the 
possibility of their presence near the surface even when their attractive 
interaction is not strong enough for them to exist in the bulk 
\cite{matsumoto,fogelstrom}.

The presence of the 
secondary order parameter near a surface is manifested in tunneling spectra 
as a splitting of the zero energy conductance peak (ZEP)
at low temperatures at zero external field and further non-linear 
splitting with increasing external field. \cite{covington}  
The field dependence of the splitting of the ZEP in 
the tunneling spectra of YBCO has been examined \cite{aprili,krupke}.
The observation
is consistent with a $d_{x^2-y^2}+is$ surface order parameter
or a $d_{x^2-y^2}+id_{xy}$ order parameter. 

In this paper we extend the BTK formula to calculate the tunneling 
conductance in a normal metal / insulator /
($d_{x^2-y^2}+is$)-wave, 
or ($d_{x^2-y^2}+id_{xy}$)-wave superconductor. 
In particular we find that in the $d_{x^2-y^2}+is$-state 
the conductance peak remains rigid at the energy of the 
subdominant ($s$) order
parameter. \cite{tanaka2,ting2} 
Besides in the $d_{x^2-y^2}+id_{xy}$ state, there is a plateau region
inside the gap due to the formation of bound states at discreet 
values of the quasiparticle trajectory 
angle $\theta$,
for all junction orientations.
Also the evolution of the tunneling conductance with temperature
depends on the nature of the subdominant order parameter.
These features can be used to distinguish between states with 
broken time-reversal symmetry.

\section{The model for the NS interface}

\begin{figure}
  \centerline{\psfig{figure=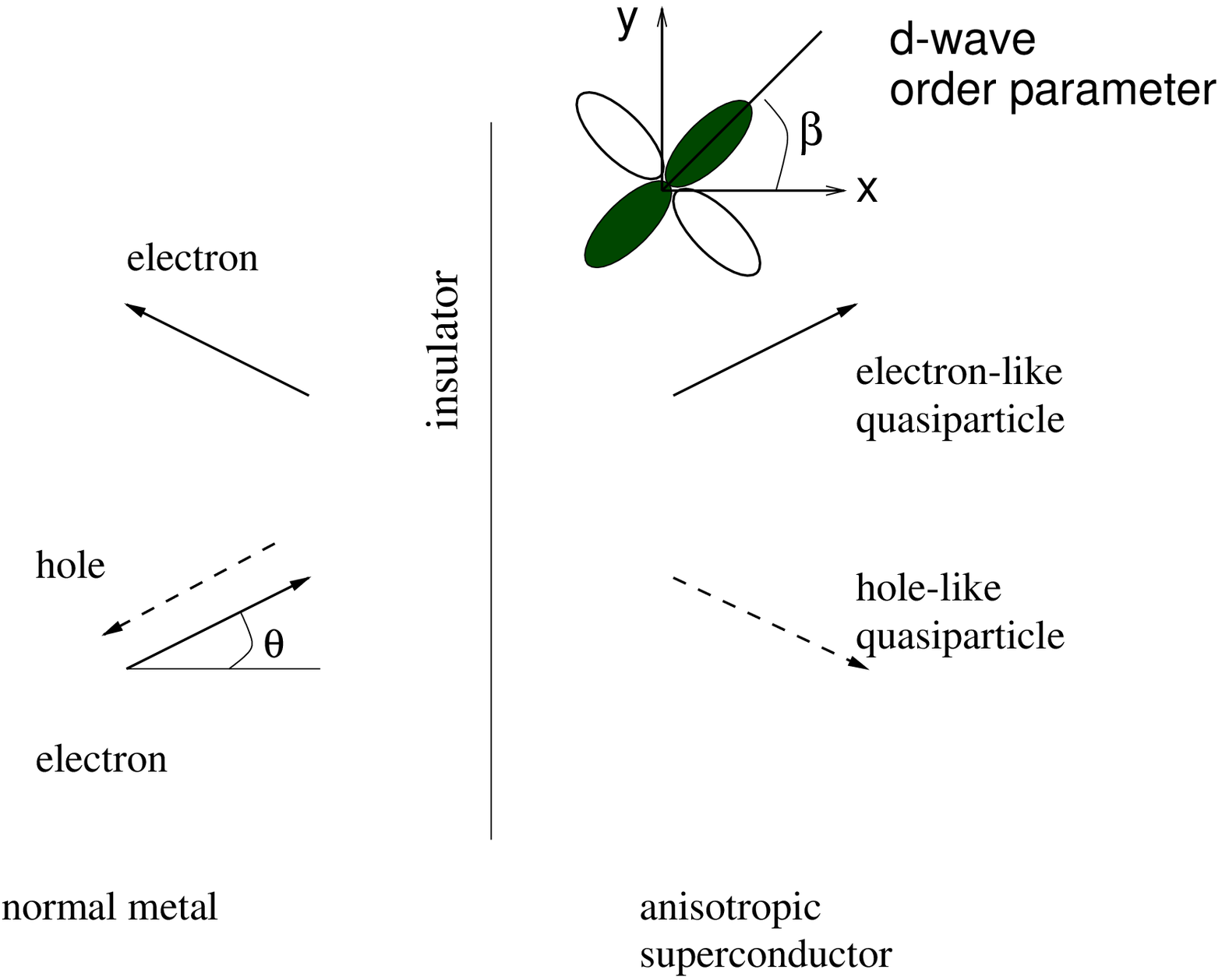,width=8.5cm,angle=0}}
  \caption{
The geometry of the normal metal / insulator / superconductor 
interface. The vertical line along the $y$-axis represents the 
insulator. 
The arrows illustrate the transmition and reflection processes at the 
interface. In this figure, $\beta$ is the angle 
between the normal to the interface and the a-axis of superconductor,
and $\theta$ is the angle of the incident electron beam and the normal.
On the top, the $d$-wave order parameter is shown.
}
  \label{fig1.fig}
\end{figure}

We consider the normal metal / insulator / superconductor 
junction shown in Fig. \ref{fig1.fig}.
We choose the 
$y$ direction to be parallel to the interface, and the $x$ direction 
to be normal to the interface.
The insulator is modeled by a delta function, located at $x=0$, of the 
form $V\delta(x)$. The temperature is fixed to $0K$.

The motion of quasiparticles in inhomogeneous superconductors 
is described by the BdG 
equations
\begin{equation}
{
\begin{array}{ll}
{\cal H}_e({\bbox r})u({\bbox r})+\int d{\bbox r'}\Delta({\bbox s},{\bbox x})v({\bbox r'}) & = Eu({\bbox r}) \\
\int d{\bbox r'}\Delta^{\ast}({\bbox s},{\bbox x})u({\bbox r'})-{\cal H}_e^{\ast}({\bbox r})v({\bbox r}) & = Ev({\bbox r})
\end{array},~~~\label{bdg}
}
\end{equation}
where the single-particle Hamiltonian is given by ${\cal H}_e({\bbox r})=
-\hbar^2\bigtriangledown_{\bbox r}^2/2m_e+V({\bbox r})-E_F$, $E$ is the energy 
measured from the Fermi energy $E_F$.
$\Delta({\bbox s},{\bbox x})$ is the pair potential, after a transformation 
from the position coordinates ${\bbox r},{\bbox r'}$ to the center of mass 
coordinate ${\bbox x}=({\bbox r}+{\bbox r'})/2$ 
and the relative vector ${\bbox s}={\bbox r}-{\bbox r'}$. 
After Fourier transformation the pair potential depends on the 
related wave vector ${\bbox k}$ and ${\bbox x}$. In the weak coupling limit ${\bbox k}$ is 
fixed on the Fermi surface ($|{\bbox k}|=k_F$), and only its direction $\theta$ is 
a variable. Also we neglect any spatial variation near the interface,
e.g. the pair potential does not depend on ${\bbox x}$. 
The pair 
potential has the form:
\begin{equation}
  \Delta({\bbox x},\theta) = \left\{ 
    \begin{array}{ll}
      0, & x < 0 \\ 
      \Delta(\theta), & x > 0  
    \end{array},~~~\label{delta}
\right.
\end{equation}
where $\theta$ is the angle of the quasiparticle trajectory
measured from $x$-axis.
When a beam of electrons is incident from the normal metal 
to the insulator, with an angle $\theta$, the general solution 
of Eqs. (\ref{bdg}), is the two component wave function,
which for $x<0$ is written
\begin{equation}
\Psi_I=
\left(
\begin{array}{ll}
  1 \\
  0 
\end{array}
\right)
e^{iq_ex\cos\theta}+a
\left(
\begin{array}{ll}
  0 \\
  1 
\end{array}
\right)
e^{iq_hx\cos\theta}+b
\left(
\begin{array}{ll}
  1 \\
  0 
\end{array}
\right)
e^{-iq_ex\cos\theta},
~~~\label{x_}
\end{equation}
while for $x>0$, the solution is 

\begin{equation}
\Psi_{II}=
c \left(
\begin{array}{ll}
  u_{+}\phi_{+} \\
  v_{+} 
\end{array}
\right)
e^{ik_ex\cos\theta}+d
 \left(
\begin{array}{ll}
  v_{-}\phi_{-} \\
  u_{-} 
\end{array}
\right)
e^{-ik_hx\cos\theta},
~~~\label{x+}
\end{equation}
where $a,b$, are the amplitudes for Andreev and normal reflection, 
and $c,d$ are the amplitudes for transmission into the superconductor 
as electron-like and hole-like quasiparticles respectively. 
In the following we assume that $q_e \approx q_h \approx k_e \approx k_h 
\approx k_F$. The latter approximation is valid within 
the BCS weak coupling theory.
The BCS coherence factors are given by 
\begin{equation}
u_{\pm}^2=[1+
      \sqrt{E^2-|\Delta_{\pm}(\theta)|^2}/E]/2,
\end{equation}
and
\begin{equation}
v_{\pm}^2=[1-
      \sqrt{E^2-|\Delta_{\pm}(\theta)|^2}/E]/2,
\end{equation}
The internal phase coming from the energy gap is given by
$\phi_{\pm} =[
\Delta_{\pm}(\theta)/|\Delta_{\pm}(\theta)|]$,
where $\Delta_{+}(\theta)=\Delta(\theta)$
($\Delta_{\_}(\theta)=\Delta(\pi- \theta)$), is the 
pair potential experienced by the transmitted electron-like 
(hole-like) quasiparticle respectively.
Using the matching conditions of the wave function at $x=0$,
$\Psi_I(0)=\Psi_{II}(0)$ and 
$\Psi_{II}'(0)-\Psi_{I}'(0)=(2mV/\hbar^2)\Psi_I(0)$, 
the magnitude of the Andreev and normal reflection $R_a=|a|^2$ and 
$R_b=|b|^2$,
are obtained as \cite{tanaka2}
\begin{equation}
R_a=\frac{\sigma_N^2|n_{+}|^2}
     {|1+(\sigma_N-1)n_{+}n_{-}
      \phi_{-}\phi_{+}^{\ast}|^2}
,~~~\label{ra}
\end{equation}

\begin{equation}
R_b=\frac{(1-\sigma_N)|1-n_{+}n_{-} 
     \phi_{-}\phi_{+}^{\ast}|^2}
     {|1+(\sigma_N-1)n_{+}n_{-}
      \phi_{-}\phi_{+}^{\ast}|^2}
,~~~\label{rb}
\end{equation}
where $n_{\pm}=v_{\pm}/u_{\pm}$.
The tunneling conductance, normalized by that in the normal 
state is given by \cite{blonder}
\begin{equation}
\sigma(E)=\frac{\int_{-\pi/2}^{\pi/2}d\theta
 \overline{\sigma}_s(E,\theta)}
{\int_{-\pi/2}^{\pi/2}d\theta\sigma_N }
,~~~\label{ss}
\end{equation}
according to the BTK formula the conductance of the junction, 
 $\overline{\sigma}_s(E,\theta)$, is expressed in terms of the 
probability amplitudes $a$, and $b$:
 $\overline{\sigma}_s(E,\theta)=1+R_a-R_b$. 
The transparency of the junction $\sigma_N$ is connected to 
the barrier height $V$ by the relation 
\begin{equation}
\sigma_N=\frac{4 \cos^2\theta}{Z^2+4 \cos^2\theta}
,~~~\label{sn}
\end{equation}
where $Z=2 m V / \hbar^2 k_F$, denotes the strength of the barrier. 
In the $Z=0$(large $\sigma_N$) limit the interface is 
regarded as a weak link, showing metallic behavior 
while for large $Z(\sigma_N=0)$
values the interface becomes insulating.

We consider the following cases:
 
a) In case of $d_{x^2-y^2}$-wave superconductor
\begin{equation}
\Delta(\theta)=
\Delta_1(T)\cos[2(\theta - \beta)]
,~~~\label{deltad}
\end{equation}
where $\beta$ denotes the angle between the normal to the interface 
and the $x$-axis of the crystal.
The temperature dependence of the gap follows the usual 
BCS relation
$\Delta_1(T)=\Delta_d \sqrt{1-T/T_d}$,
where $T_d$ is the transition temperature.

b) In the ($d_{x^2-y^2}+is$)-wave case
\begin{equation}
\Delta(\theta)=
\Delta_1(T)\cos[2(\theta - \beta)] + i \Delta_2(T)
,~~~\label{deltadis}
\end{equation}
where $\Delta_2(T)=\Delta_s \sqrt{1-T/T_s}$,
and $T_s$ is the transition temperature for the 
$s$-wave component.

c) In the ($d_{x^2-y^2}+id_{xy}$)-wave case
\begin{equation}
\Delta(\theta)=
\Delta_1(T)\cos[2(\theta - \beta)] + i \Delta_2(T)\sin[2(\theta - \beta)]
,~~~\label{deltadid}
\end{equation}
where the angular form of the secondary component is obtained by 
the substitution of $\beta$ in the $d_{x^2-y^2}$-wave order 
parameter by $\beta+\pi /4$. 
$\Delta_2(T)=\Delta_{d_{xy}} \sqrt{1-T/T_{d_{xy}}}$, follows the
BCS relation, and $T_{d_{xy}}$ is the transition temperature for the 
${d_{xy}}$-wave component.

\section{Tunneling conductance characteristics} 

\begin{figure}
  \centerline{\psfig{figure=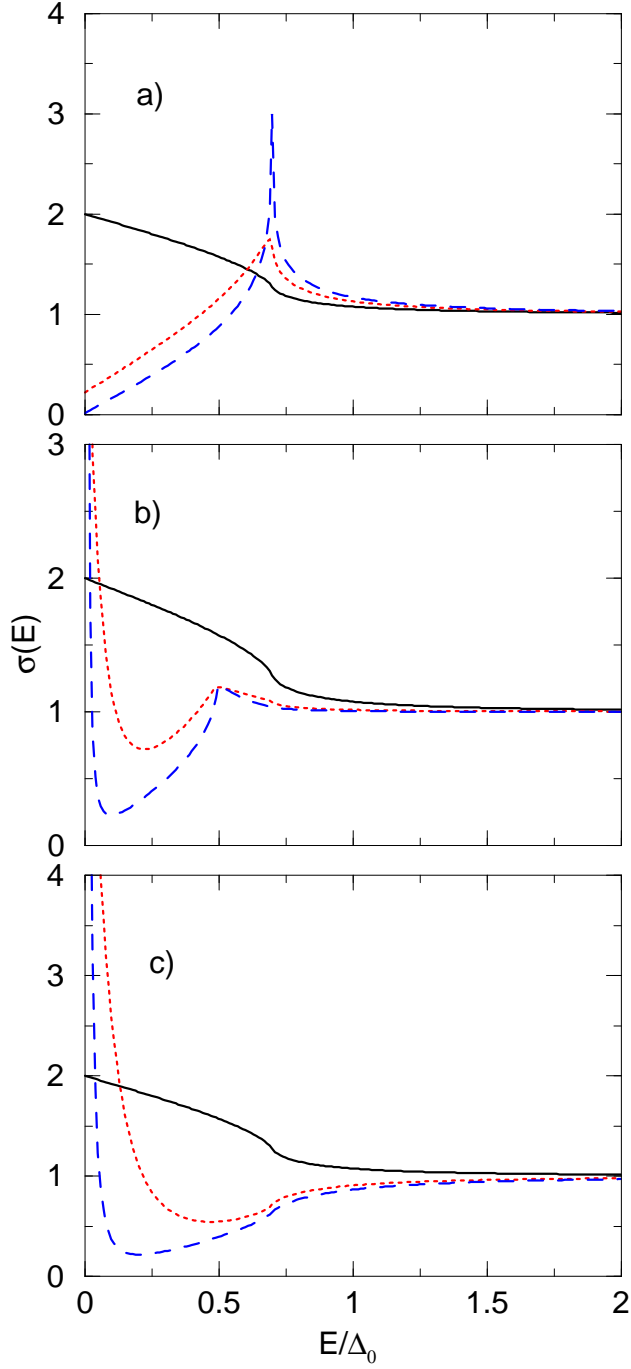,width=8.5cm,angle=0}}
  \caption{
Normalized tunneling conductance $\sigma(E)$ as a function of $E/\Delta_0$
for $Z=0$ (solid line), $Z=2.5$ (dotted line), $Z=10$ (dashed line),
for different orientations (a) $\beta=0$, (b) $\pi/8$, (c) $\pi/4$.
The pairing
symmetry of the superconductor is
$d_{x^2-y^2}$, $\Delta_d=0.7\Delta_0$. The temperature is $T=0$.
}
  \label{d.fig}
\end{figure}

\begin{figure}
  \centerline{\psfig{figure=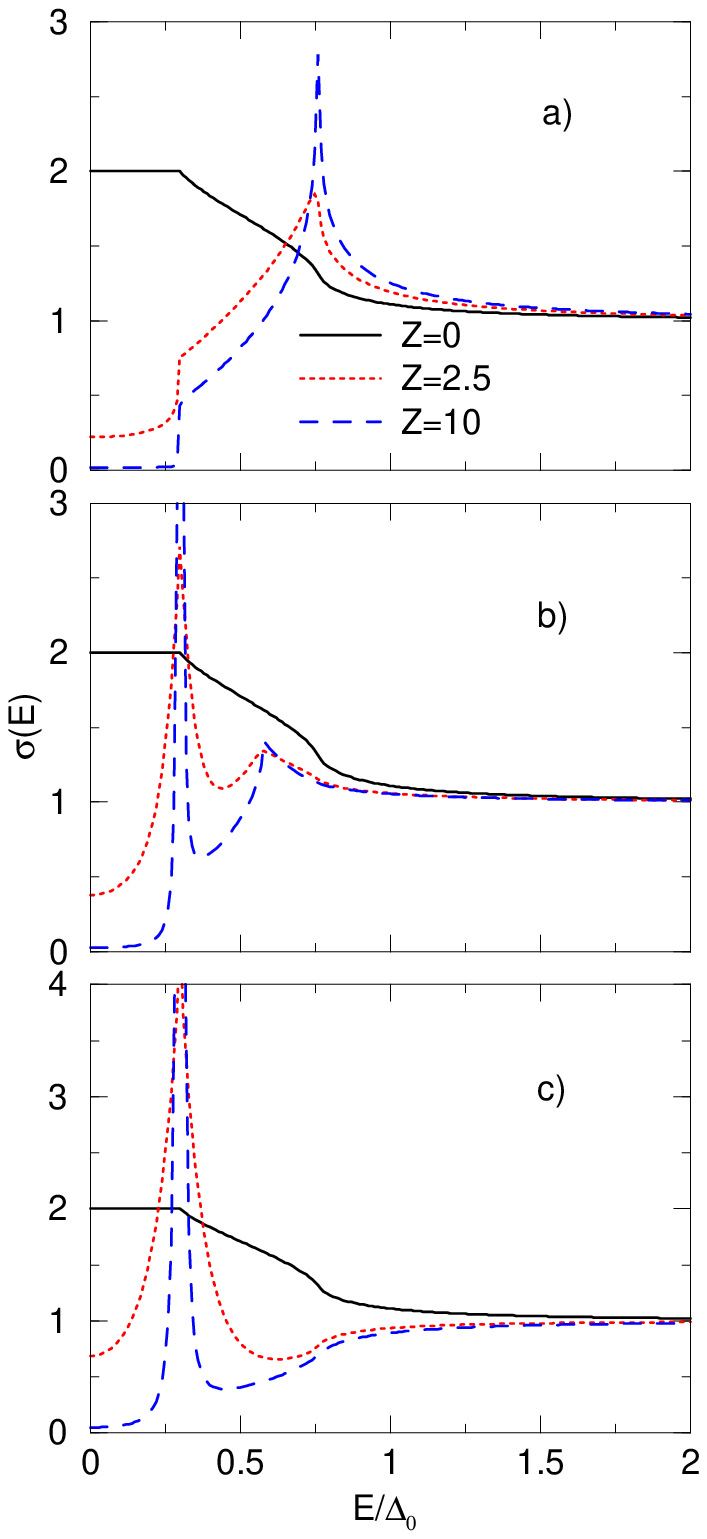,width=8.5cm,angle=0}}
  \caption{
The same as in Fig. 2. The pairing
symmetry of the superconductor is
$d_{x^2-y^2}+is$, $\Delta_d=0.7\Delta_0$, $\Delta_s=0.3\Delta_0$.
}
  \label{dis.fig}
\end{figure}

\begin{figure}
  \centerline{\psfig{figure=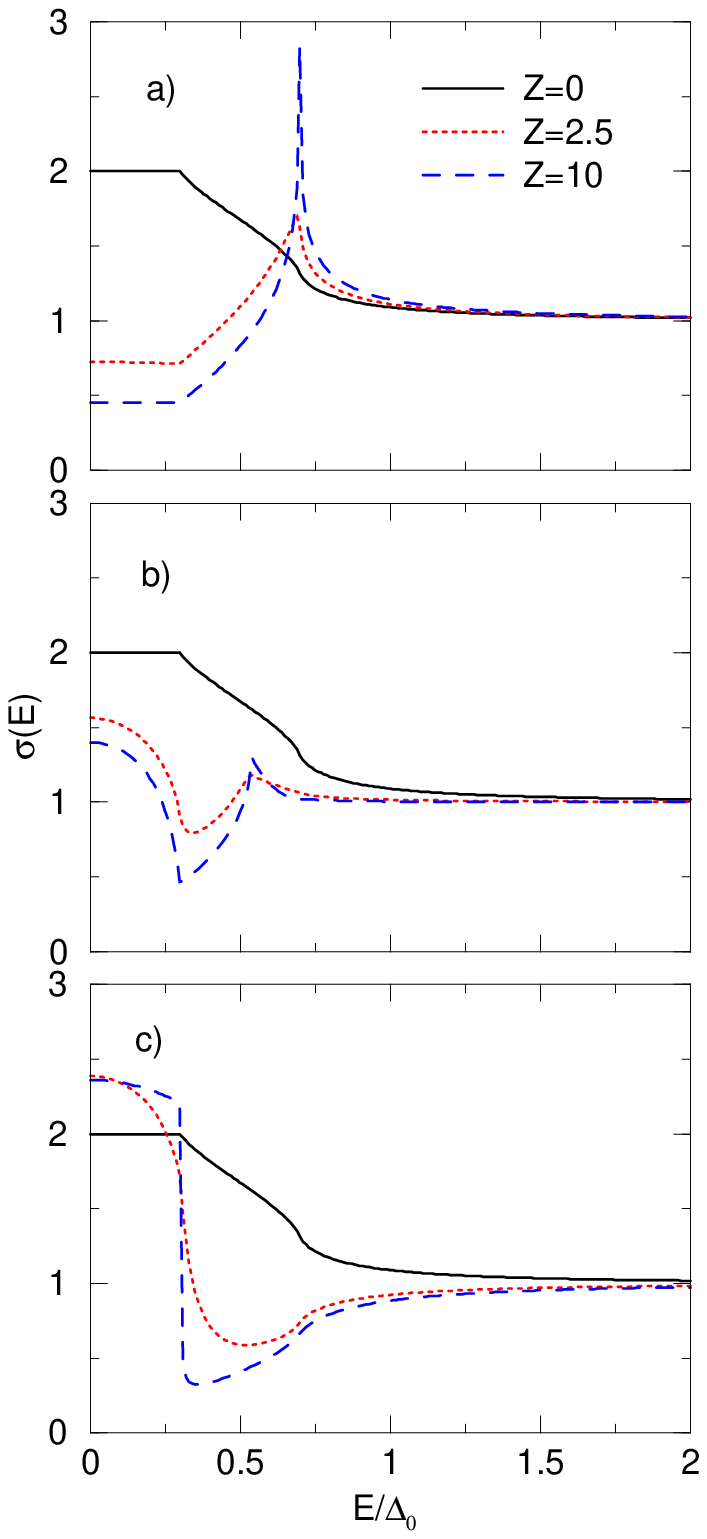,width=8.5cm,angle=0}}
  \caption{
The same as in Fig. 2. The pairing
symmetry of the superconductor is
$d_{x^2-y^2}+id_{xy}$, $\Delta_d=0.7\Delta_0$, $\Delta_{d_{xy}}=0.3\Delta_0$.
}
  \label{did.fig}
\end{figure}

In Figs. 2-4 we plot the tunneling conductance $\sigma(E)$
as a function of $E/ \Delta_0$
for various values of $Z$, for different orientations (a) $\beta=0$, 
(b) $\pi/8$,
(c) $\pi/4$. The pairing
symmetry of the superconductor is
$d_{x^2-y^2}$-wave, with $\Delta_d=0.7\Delta_0$, in Fig. \ref{d.fig},
($d_{x^2-y^2}+is$)-wave, with $\Delta_d=0.7\Delta_0$, $\Delta_s=0.3\Delta_0$, 
in Fig. \ref{dis.fig}),
($d_{x^2-y^2}+id_{xy}$)-wave, with $\Delta_d=0.7\Delta_0$, 
$\Delta_{d_{xy}}=0.3\Delta_0$
in Fig. \ref{did.fig}.
It is clear from these figures that 
the peaks are narrowed by the increase of $Z$.
In this section the temperature is fixed to $0 K$.

For $\beta=0$ e.g. when the lobes of the dominant $d$-wave component 
point towards the junction interface the position of the conductance peak, is 
near the energy gap $\Delta_d$, in all the above pairing symmetries.
This peak is mainly effected from the bulk density of states. 

For $\beta \neq 0$ another peak exists in the conductance spectra, 
for the $d_{x^2-y^2}$-wave, ($d_{x^2-y^2}+is$)-wave cases,
but its physical origin is different than that found near $\Delta_d$.
For the $d$-wave case this peak exists at $E=0$ for all the non zero values 
of $\beta$, due to the different sign of the pair potential
that the transmitted quasiparticles feel. However,
the height of the conductance peak (ZEH) depends on the orientation 
angle $\beta$.
For a given angle $\beta$ the ZEH is proportional 
to the range of $\theta$ angles for which sign change 
occurs. This is seen in Fig. \ref{d.fig} (c) for $\beta=\pi /4$ 
where the ZEH is maximum 
since for this orientation the transmitted quasiparticles feel 
different sign of the pair potential for all angles 
$-\pi /2 < \theta < \pi /2$. 
On the other hand for $\beta=\pi /8$ in  Fig. \ref{d.fig} (b) the 
range of angles is reduced and the ZEH takes a lower value. 

For the ($d_{x^2-y^2}+is$)-wave case in Fig. \ref{dis.fig} 
the position of the conductance peak is shifted to the 
energy $E=\Delta_s$, for all values of $\beta$. For each value 
of $\beta$,
its height depends on the range of $\theta$ angles 
where the transmitted quasiparticles feel different sign 
of the pair potential. For $\beta=\pi /4$ the conductance peak 
as seen in Fig. \ref{dis.fig} (c) 
at $E=\Delta_s$, has its maximum value since the transmitted quasiparticles 
feel the sign change of the pairing potential for all angles 
$\theta$. This range is reduced for other orientations and 
for $\beta=0$ it goes to zero, as we can see in Fig. \ref{dis.fig} (a).
Also a subgap opens within the conductance spectra due to the 
imaginary $s$-wave component.
Within the subgap in the tunneling limit, 
the tunneling conductance is zero, $\sigma(E)=0$ while
in the metallic limit ($Z=0$), $\sigma(E)=2$ independently 
from the orientation, as in the 
$s$-wave case. In the $Z=0$ case the normal reflection coefficient 
is zero while the Andreev reflection coefficient is unit. In this case the 
charge transport into the superconductor is twice as large 
as in the normal state, for energies within the subgap region.

In the $d_{x^2-y^2}+id_{xy}$ case, seen in Fig. \ref{did.fig} 
the tunneling conductance has residual values within the gap 
for all orientations $\beta$.
In particular for $\beta=0$, as seen in Fig. \ref{did.fig} (a) in the 
tunneling limit, the conductance $\sigma(E=0)$, at $E=0$
has a non zero value contrary to the ($d_{x^2-y^2}+is$)-wave 
case where it is zero. In the $d_{x^2-y^2}+is$ case
for $\beta=0$, there is no angle $\theta$ for which the 
transmitted quasiparticles to experience the sign change of the 
pair potential, and the tunneling conductance goes to zero. 
This is different for the $d_{x^2-y^2}+id_{xy}$ 
where for $\beta=0$ the
transmitted quasiparticles feel 
the sign difference due to the  secondary order parameter $d_{xy}$. 

\begin{figure}
  \centerline{\psfig{figure=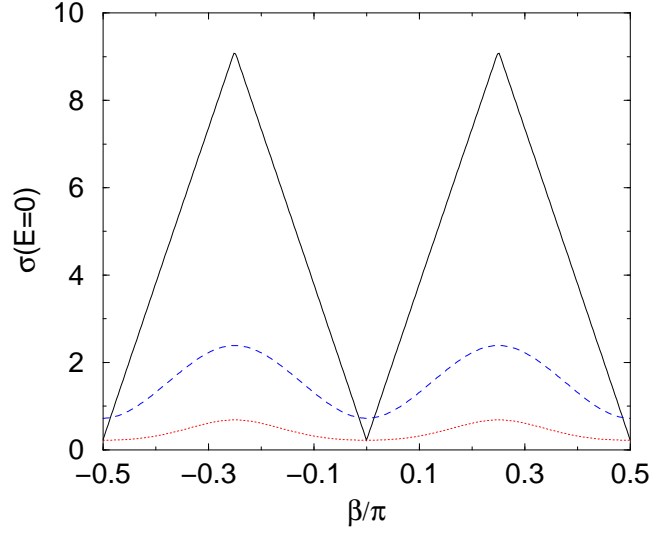,width=8.5cm,angle=0}}
  \caption{
Normalized tunneling conductance $\sigma$ for $E=0$ 
as a function of $\beta$
for $Z=2.5$, and $T=0$. The pairing 
symmetry of the superconductor is 
$d_{x^2-y^2}$ (solid line), $\Delta_d=0.7\Delta_0$, 
$d_{x^2-y^2}+is$ (dotted line), $\Delta_d=0.7\Delta_0$, $\Delta_s=0.3\Delta_0$,
and $d_{x^2-y^2}+id_{xy}$ (dashed line), $\Delta_d=0.7\Delta_0$, 
$\Delta_{d_{xy}}=0.3\Delta_0$.
}
  \label{sE0.fig}
\end{figure}

Also the zero energy conductance height evolves very differently 
with the orientation 
of the superconductor, for the three pairing symmetries. This is seen in
Fig. \ref{sE0.fig}  where the dependence of zero-energy 
conductance height on $\beta$
is plotted, for $Z=2.5$, for the three pairing symmetries. 
It is seen that for the ($d_{x^2-y^2}+id_{xy}$)-wave case (dashed line), for 
$\beta$ close to $\pi /4$, 
the height representing the plateau like feature seen in Fig. \ref{did.fig}
is enhanced. 
Besides for angles close to zero, 
the height 
of the ZEP for the ($d_{x^2-y^2}+id_{xy}$)-wave case is reduced, but 
is not zero. Note that the height for
$\beta=0$, remains finite even in the 
large $Z$-limit, while the height in the ($d_{x^2-y^2}+is$)-wave case 
goes to zero. 

\section{Bound state energies}
\begin{figure}
  \centerline{\psfig{figure=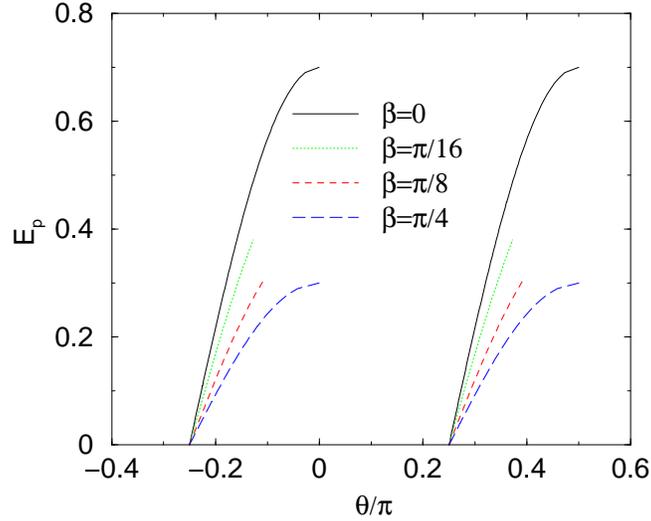,width=8.5cm,angle=0}}
  \caption{
Bound state energy $E_p$, for $T=0$, versus the quasiparticle angle 
$\theta$ for different orientations $\beta=0, \pi /16,
\pi /8, \pi /4$.
The pairing 
symmetry of the superconductor is 
$d_{x^2-y^2}+id_{xy}$ with $\Delta_d=0.7\Delta_0$, $\Delta_{d_{xy}}=0.3\Delta_0$.
}
  \label{didbs.fig}
\end{figure}

\begin{figure}
  \centerline{\psfig{figure=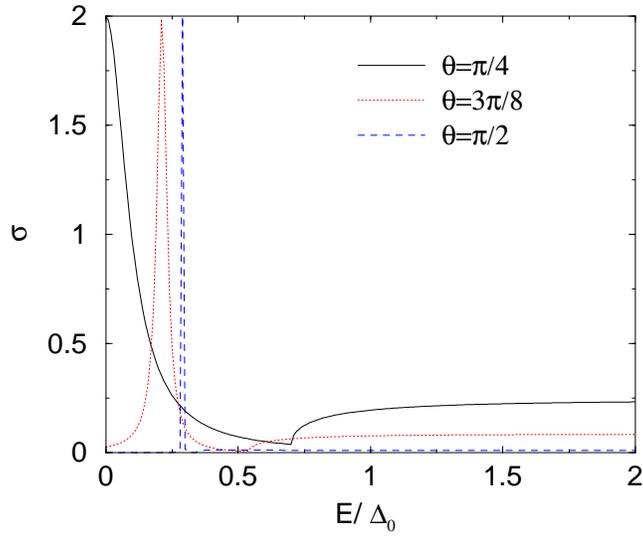,width=8.5cm,angle=0}}
  \caption{
Conductance $\overline{\sigma}_s (E,\theta)$ for $Z=2.5$, and $T=0$, as a 
function of $E$ for fixed angle $\beta=\pi /4$, at 
different angles $\theta=\pi /4, 3\pi /8, \pi /2$, for 
which a bound state is formed at a different value 
of $E$.
The pairing symmetry of the superconductor is 
$d_{x^2-y^2}+id_{xy}$ with $\Delta_d=0.7\Delta_0$, 
$\Delta_{d_{xy}}=0.3\Delta_0$.
}
  \label{overline.fig}
\end{figure}

These features are explained if we calculate the energy of the 
midgap state, which is given for large $Z$ by the value in which 
the denominator of Eq. (\ref{ra},\ref{rb}) vanishes. The equation giving the 
energy peak level is written as \cite{tanaka2}
\begin{equation}
     \phi_{-} \phi_{+}^{\ast}n_{+}n_{-}|_{E=E_p}=1.0
.~~~\label{midgap}
\end{equation}
In the $d_{x^2-y^2}$-wave case, for a given angle $\beta$, 
this equation has solution
$E=0$,
for a finite range
of angles $\theta$. 
For $\beta=\pi /4$ the solution is $E=0$
for $-\pi /2 < \theta < \pi /2$, since 
$n_{+}n_{-}|_{E=0}=-1$, and also the transmitted quasiparticles
feel different sign of the pair potential i.e. 
$\phi_{-}\phi_{+}^{\ast}|_{E=0}=-1$.
In the ($d_{x^2-y^2}+is$)-wave case for $\beta=\pi /4$, the 
solution is $E=\Delta_s$ in the $\theta$ interval $[0:\pi /2]$. 
In this case the $n_{+}, n_{-}$ and the internal phases are 
varied in a way that Eq. \ref{midgap} is satisfied for 
$E=\Delta_s$ and a midgap state is formed.
When a midgap state exists the tunneling conductance 
$\overline{\sigma}_s(E,\theta)$ is equal to $2$ for all $\theta$ 
and the peek in $\sigma(E)$
seen in Fig. \ref{d.fig},\ref{dis.fig}, is due to the normal state
conductance $\sigma_N$ in Eq. \ref{sn}, which depends inversely on the $Z^2$
for large $Z$. For intermediate angles $\beta$,
the peak height of the tunneling conductance $\sigma(E)$ is
proportional to $8\beta$, for $0<\beta<\pi /4$, \cite{tanaka1} and 
for $\beta=0$ the range of $\theta$ angles for which Eq. \ref{midgap} 
has solutions collapses to zero in 
both symmetry states, and no bound states are formed. Then 
$\sigma(E)$ goes to zero as $1/Z^2$ and there is no conductance peak.
For energies different than the bound state energy $E_p$, for 
large $Z$, $\overline{\sigma}_s(E,\theta)$ is inversely proportional 
to $Z^2$ as the $\sigma_N$ is and the tunneling conductance 
has a constant value as we can see in Fig. \ref{d.fig}, for $E>0$. 
In the ($d_{x^2-y^2}+id_{xy}$)-wave case 
for fixed $\beta$ the solutions of \ref{midgap} depend 
both on $E$, and $\theta$, as seen in Fig. \ref{didbs.fig}, where the 
bound state energy $E_p$ is plotted for $\beta=0, \pi /16, \pi /8, \pi /4$, 
as a function of $\theta$. 
In this case the midgap state for a given 
$\beta$ is formed for a pair of angles $\theta$, for 
energies within the gap. 
This observation can be used to explain the residual values 
of the tunneling conductance 
within the gap, 
seen in Fig. \ref{did.fig} as follows. 
When a bound state is formed the conductance 
$\overline{\sigma}_s(E,\theta)$ is equal to $2$ exactly 
at the bound state energy for the two discreet values of $\theta$ 
and the 
peak in the $\sigma(E)$  should be proportional to the $Z^2$ for 
large $Z$ for these values of $\theta$. 
For the rest of the quasiparticle trajectory angles $\theta$
the tunneling conductance
$\sigma(E)$, has a constant value. 
Thus the height for a given energy, and angle $\beta$ is 
determined from the interplay of two competitive factors, i.e.,
the bound state energy formed at a couple of $\theta$ angles which
gives a contribution proportional to $Z^2$, and the rest of 
$\theta$ angles which give a constant value contribution 
independent of $Z$.
Also the steps in $\theta$, 
in evaluating the integral 
in Eq. \ref{ss} are very much crucial since the calculation of the tunneling 
conductance has to be performed exactly
at the bound state energy. If this is not the case then
the peak due to the bound states 
in the tunneling conductance would have a smaller value which 
would depends
on $Z$ in general.
We conclude that in the $d_{x^2-y^2}+id_{xy}$ 
the discreet values of the quasiparticle trajectory angle 
$\theta$, over which a bound state is formed, compared 
to the interval of $\theta$ angles in the other two pairing 
states explains the reduced height of the tunneling conductance
within the gap. 
However if we calculate the conductance $\overline{\sigma}_s (E,\theta)$, 
for the $d_{x^2-y^2}+id_{xy}$ at a given $\beta$ for a value of 
$\theta$ for which bound state occurs then the conductance
should develop a peak at the bound state energy, where
$\overline{\sigma}_s (E,\theta)$ is equal to $2$. For the rest 
of the energies, 
$\overline{\sigma}_s (E,\theta)$, goes to zero as $1/Z^2$.
This is seen in Fig. \ref{overline.fig} where the 
conductance $\overline{\sigma}_s (E,\theta)$ for $Z=2.5$ is plotted for 
fixed $\beta=\pi /4$ as a function of the energy $E$, for 
different values of the angle $\theta=\pi /4, 3\pi /8, \pi/2$
for which bound state occurs. 
We see that for $\theta=\pi /4$ the peak is at $E=0$. 
However as we change the angle $\theta$ towards $\pi /2$ the peak level 
moves from $E=0$ to $E=\Delta_{d_{xy}}$.

The occurence of residual density of states in the 
($d_{x^2-y^2}+id_{xy}$)-wave case is unaffected by the calculation 
of $\sigma(E)$ including the self-consistency of the order parameter.
\cite{tanaka4}
In this calculation an enhancement appears at $E=\Delta_{d_{xy}}$, 
for $\beta=\pi /4$. In our calculation 
we also observed a similar enhancement at $E=\Delta_{d_{xy}}$ 
when the definition 
\begin{equation}
\sigma(E)=\frac{\int_{-\pi/2}^{\pi/2}d\theta
 \overline{\sigma}_s(E,\theta)\cos \theta}
{\int_{-\pi/2}^{\pi/2}d\theta\sigma_N \cos \theta}
,~~~\label{ssnew}
\end{equation}
was used for the calculation of the tunneling conductance. 
The $\cos \theta$ factor was included in the integration formula 
to calculate the $x$-component of the tunneling spectra. 
Within this definition the bound state at $\theta=0$ 
contributes more (due to the $\cos \theta$ factor) that the 
bound state at $\theta$ close to $\pi /4$. As seen in figure \ref{didbs.fig}
the bound state at $\theta=0$ corresponds to energy $E=\Delta_{d_{xy}}$
causing the peak in $\sigma(E)$ at $E=\Delta_{d_{xy}}$. 
Also in a self consistent calculation the bound state 
at $(\theta=\pi/4, E=0)$ contribute less in $\sigma(E)$ that that at 
$(\theta=\pi/4, E=\Delta_{d_{xy}})$ due to the depletion near 
the interface. In any case the peak near $E=\Delta_{d_{xy}}$ in 
($d_{x^2-y^2}+id_{xy}$)-wave pairing state is much more suppressed 
that that at $E=\Delta_s$ in ($d_{x^2-y^2}+id$)-wave state.

The angular dependence of the bound state energy for fixed 
boundary orientation at the $xy$ plane can be used to identify 
the ($d_{x^2-y^2}+id_{xy}$)-wave pairing state. The method we 
propose here is the two point spectroscopy described by 
Benistant et. al. \cite{benistant}. They measured the reflected 
hole distribution along the boundary $y$-direction, when electrons  
are injected with certain distribution $P(\phi)$, through a 
point contact, at $y=0$, into a normal metal of thickness $d$ attached 
to an $s$-wave superconductor. 
The presence of a magnetic field parallel to the $z$-axis 
deflects the trajectories of the electrons
and leads to an asymmetric distribution of angles of incidence 
in the normal metal / superconductor interface. 
Also the magnetic field focuses the 
reflected holes into a second point contact which acts as 
a hole collector. Moving the second point contact around the 
first one or using several point contacts along the direction 
parallel to the interface  we are able to measure 
the intensity of the Andreev reflected holes as a function of the 
$y$ direction. 
In the $s$-wave case one observes a  
single peak called 'focusing peak' at $y=y_0$ from the injection 
point at $y=0$, since the Andreev reflected probability amplitudes 
are independent of the injection angle. 
For the ($d_{x^2-y^2}+id_{xy}$)-wave
case the bound state energies,
for which the reflection coefficient 
is equal to one, occurs at angles $\theta_1<0, \theta_2>0$, for a given 
boundary orientation and large barrier strength $Z$. These bound states 
will give rise to a second peak in the hole distribution, 
at a different position, besides the one due to the focusing. 
The presence of the magnetic field leads to 
an asymmetric distribution of angles of incidence in the interface 
and the trajectory which corresponds to bound state at $\theta_1$ 
has shorter path from that at $\theta_2$ 
and the corresponding injected electrons have
smaller angle $\phi$.
If the angular distribution probability $P(\phi)$ of the injected 
particles is peaked at small injection angles, this will lead 
to larger contribution to the secondary peak from the bound state 
at $\theta_1$ then that at $\theta_2$
This new peak would be observed for all energies for which bound 
states exist in a ($d_{x^2-y^2}+id_{xy}$)-wave superconductor. 
In the case of a $d_{x^2-y^2}$-wave superconductor, 
and a ($d_{x^2-y^2}+is$)-wave superconductor
the resonance exist only for $E=0$, $E=\Delta_s$ correspondingly.
Experiments of this kind require high quality normal conducting 
crystal and point contacts for the electron injection. 
Any voltage drop has to occur at the point contact for the 
electrons to move ballistically in the normal metal. 
Similar procedure has been proposed from Honerkamp and Sigrist 
\cite{honerkamp} to discriminate between unitary and nonunitary 
triplet states for the superconductor Sr$_2$RuO$_4$.

\section{Temperature dependence of the tunneling spectra}
At finite temperatures the tunneling conductance is calculated 
from the relation \cite{tinkham}
\begin{equation}
\sigma(eV)=\int_{-\infty}^{\infty}dE
\left[ -\frac{\partial f(E+eV)}{\partial E}
\right]
\frac{\int_{-\pi/2}^{\pi/2}d\theta
\overline{\sigma}_s(E,\theta)}
{\int_{-\pi/2}^{\pi/2}d\theta
\sigma_N}
,~~~\label{sev}
\end{equation}
$eV$ is the electron energy and $f(E)$ is the Fermi function 
$f(E)=1/(e^{\beta E}+1)$, $\beta=1/k_B T$. 
In the case of a two component order parameter, 
we assume that below a surface transition temperature a 
subdominant order parameter can develop which breaks spontaneously 
the time reversal symmetry. 
Its amplitude is bellow the value for the 
formation of spontaneously broken time reversal symmetry 
state in the bulk. \cite{fogelstrom}
The temperature dependence of the pair potential amplitude 
is assumed to obey the usual BCS relation.  
As a consequence under the coexistence of the secondary component 
the $T_c$'s for the dominant $d$-wave $T_d$ and the subdominant 
$s$($d_{xy}$) components, $T_s$($T_{d_{xy}}$)
directly correspond to the amplitude of the attractive interaction 
in each case. 

\begin{figure}
  \centerline{\psfig{figure=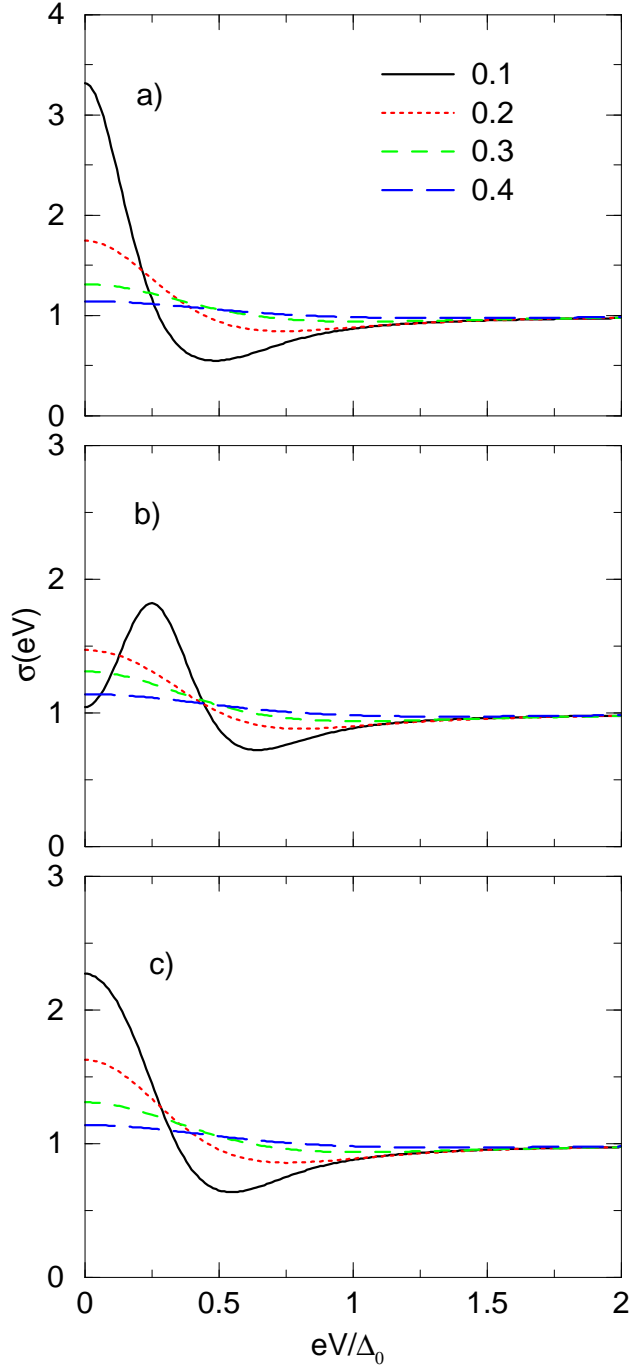,width=8.5cm,angle=0}}
  \caption{
Normalized tunneling conductance $\sigma$ versus the applied 
voltage $eV$, for different temperatures $T/T_d=0.1,0.2,0.3,0.4$. 
The barrier strength is $Z=10$, and the junction orientation 
is fixed to $\beta=\pi /4$. The pairing 
symmetry of the superconductor is 
$d_{x^2-y^2}$ in a), 
$d_{x^2-y^2}+is$ with $T_s=0.3T_d$ in b),
and $d_{x^2-y^2}+id_{xy}$ with $T_{d_{xy}}=0.3T_d$ in c).
}
  \label{temp.fig}
\end{figure}

Fig. \ref{temp.fig} shows the tunneling conductance $\sigma(eV)$ 
for different temperatures $T/T_d=0.1,0.2,0.3,0.4$, in the large barrier 
strength limit $Z=10$, $\beta=\pi/4$. 
The pairing
symmetry of the superconductor is
$d_{x^2-y^2}$-wave in Fig. \ref{temp.fig} (a),
($d_{x^2-y^2}+is$)-wave, with $T_s=0.3T_d$,
in Fig. \ref{temp.fig} (b),
($d_{x^2-y^2}+id_{xy}$)-wave, with $T_{d_{xy}}=0.3T_d$
in Fig. \ref{temp.fig} (c).
It is seen that due to the thermal occupation of states contributing 
to the tunneling current, the peaks are getting broadened as 
the temperature increases. 
In the $d_{x^2-y^2}$-wave case, as seen in Fig. \ref{temp.fig} (a)
the ZEP is suppressed when the temperature increases 
and disappears almost at the critical temperature. 
This feature of the calculated spectra is consistent 
with the experimental results of YBa$_2$Cu$_3$O$_{7-\delta}$ 
observed by low-temperature scanning tunneling spectroscopy.   
\cite{tanaka3} 
The evolution of the conductance spectra with temperature 
is qualitatively similar with the calculation including 
the self-consistency. \cite{barash} 
On the other hand in the ($d_{x^2-y^2}+is$)-wave, 
seen in Fig. \ref{temp.fig} (b) the tiny subgap of the order 
of $\Delta_s=0.3\Delta_0$ at $T=0$ disappears with the increase 
of the temperature. For $T>T_s$ it follows the usual  
$d_{x^2-y^2}$-wave like dependence. 
In the ($d_{x^2-y^2}+id_{xy}$)-wave, shown 
in Fig. \ref{temp.fig} (c) the zero energy height is suppressed 
with the increase of the temperature. For $T>T_{d_{xy}}$ the temperature
dependence of the spectra for the ($d_{x^2-y^2}+id_{xy}$)-wave 
state is similar to the 
$d_{x^2-y^2}$-wave.

\begin{figure}
  \centerline{\psfig{figure=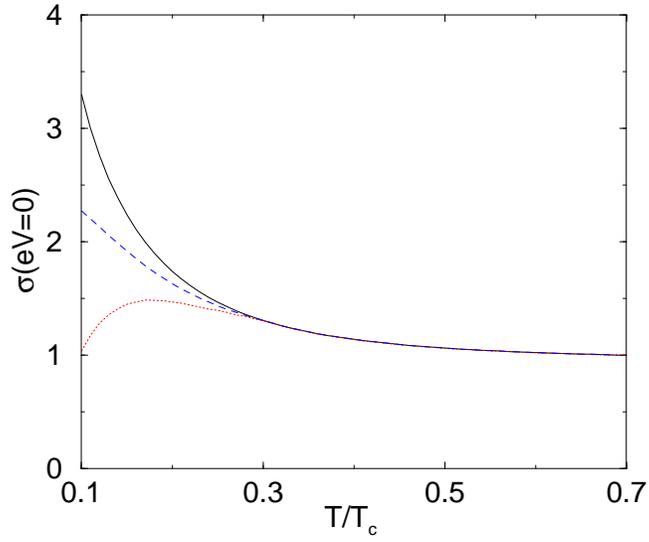,width=8.5cm,angle=0}}
  \caption{
Normalized tunneling conductance $\sigma$ for $eV=0$ 
as a function of the temperature $T/T_d$
for $Z=10$, and $\beta=\pi /4$. The pairing 
symmetry of the superconductor is 
$d_{x^2-y^2}$ (solid line), 
$d_{x^2-y^2}+is$ with $T_s=0.3T_d$ (dotted line),
and $d_{x^2-y^2}+id_{xy}$ with $T_{d_{xy}}=0.3T_d$ (dashed line).
}
  \label{sE0vsT.fig}
\end{figure}

In Fig \ref{sE0vsT.fig} we plot the ZEH as a function of 
temperature for the three pairing states. For the $d_{x^2-y^2}$-wave case
the ZEH behaves as $T^{-1}$. 
For the ($d_{x^2-y^2}+is$)-wave case the ZEH increases up to $T=0.2T_d$ 
and then decreases with increasing the temperature. 
For $T>T_s$ it follows the $d_{x^2-y^2}$-wave behavior.
The downturn of the ZEH at low temperatures in the  ($d_{x^2-y^2}+is$)-wave case
corroborates with the ZEP
splitting and has also been observed 
experimentally (see Fig.1 in Ref. \cite{covington}).
In the ($d_{x^2-y^2}+id_{xy}$)-wave case the ZEH decreases with $T$ 
in different scales for $T<T_{d_{xy}}$ and $T>T_{d_{xy}}$ 
indicating the different pairing states. 
In all cases the transition from the $d_{x^2-y^2}$-wave to the 
($d_{x^2-y^2}+is$)-wave or ($d_{x^2-y^2}+id_{xy}$)-wave is continuous.

In the metallic limit ($Z=0$) (not presented in the figure) 
the tunneling conductance at $eV=0$ decreases, as the 
temperature increases, from its zero 
temperature value $\sigma(eV)=2$, to the normal state value 
$\sigma(eV)=1$ at the transition temperature. The variation with $T$
for the $d_{x^2-y^2}+is$ ($d_{x^2-y^2}+id_{xy}$)-wave for 
$T<T_s (T_{d_{xy}}$) deviates from the $d_{x^2-y^2}$-wave behavior. 
In both cases where time reversal symmetry is broken, a change 
of slop exists in the $\sigma(eV=0)$ vs $T$ diagram, 
at the subdominant order parameter transition 
temperature. However in this case the variation with $T$ 
is similar for $d_{x^2-y^2}+is$, $d_{x^2-y^2}+id_{xy}$ and 
thus it can not be used to discriminate between the two pairing 
states. 

\section{Conclusions} 
We calculated the tunneling conductance in normal metal / insulator / 
anisotropic superconductor, using the BTK formalism. We showed that  
the conductance peak for (110) surface orientation, 
in a $d_{x^2-y^2}$-wave 
superconductor, appears in zero energy, and is shifted according to the 
amplitude of the secondary order parameter in the ($d_{x^2-y^2}+is$)-wave case.
In the ($d_{x^2-y^2}+id_{xy}$)-wave case the tunneling conductance 
has residual states 
within the energy gap.
These are due to the formation of bound states at discreet 
values of the trajectory angle $\theta$ for each boundary orientation 
angle $\beta$, for energies within the gap. 
These bound states explain both the residual states within the 
subgap and also the small height of the conductance within 
the subgap region.
The calculation of the conductance $\overline{\sigma}_s (E,\theta)$ 
for given boundary orientation at an incident angle $\theta$ for 
which bound state occurs shows an enhancement at the bound state 
energy. 
The energy dependence of the bound state on $\theta$ can be used 
within the method of electron focusing to detect the 
($d_{x^2-y^2}+id_{xy}$)-wave state. In such a case besides the 
focusing peak another peak exists in the reflected hole 
distribution spectrum for all energies of the injected electrons
less than the amplitude of the secondary order parameter. 
This peak should also be observed for the $d_{x^2-y^2}$-wave and 
($d_{x^2-y^2}+is$)-wave cases but only at the energy $E=0$, $E=\Delta_s$ 
respectively.

The zero energy conductance peak decreases as $T^{-1}$ with increasing the 
temperature and disappears almost at the transition temperature for the 
$d_{x^2-y^2}$-wave case. The temperature dependence of the ZEH 
deviates from the usual $T^{-1}$ behavior of the $d_{x^2-y^2}$, 
in the case where a subdominant surface order parameter is 
developed, for $T<T_{c1}$, where 
$T_{c1}$ is the transition temperature for the subdominant order 
parameter.
These features can be used to distinguish between time-reversal 
broken symmetry states.  

Throughout this paper the spatial variation of the dominant order parameter 
near the surface which depends on the boundary orientation 
is ignored for simplicity. As a consequence, 
since the nucleation of the 
secondary order parameter near the surface 
depends on the strength of the dominant one, the spatial 
variation of the secondary order parameter is also ignored. 
We expect more drastic changes when the orientation is $\beta=\pi/4$
where the suppression of the dominant order parameter is more significant.
However, since the features presented here are intrinsic and are 
generated by the existence of surface bound states, the essential 
results do not change qualitatively. 

Also we assumed perfectly flat interfaces in the clean limit, 
so any impurity scattering and the effect of the surface roughness 
are ignored. Generally surface roughness will lead to a statistical 
distribution of the outgoing trajectories, and will alter the 
results presented. The effect of surface roughness on the tunneling effect 
in interfaces between normal metals and superconductors with 
time reversal symmetry broken, has been studied.\cite{rainer}
It is found that in the $d_{x^2-y^2}+id_{xy}$ case 
additional bound states are formed due to the 
surface roughness.
Also in $d_{x^2-y^2}$-wave
superconductor, the ZEP may appear
even for ($100$) interfaces with surface roughness
\cite{fogelstrom}.

Also in a more realistic treatment of the problem, one has 
to take into account also the thickness of the barrier. In 
that case additional resonances are expected in the tunneling spectra
due to multiple Andreev reflections within the barrier, 
besides the ones due to the bound states. 

\section{Acknowledgments}
I would like to thank A.V. Balatsky for discussions
and for careful reading of the manuscript.
Also I wish to thank Yukio Tanaka for providing me with a list 
of his publications. 
I received partial support from the ESF programme 
FERLIN.

\newpage

\end{document}